\documentclass[compsoc, conference, a4paper, 10pt, times]{IEEEtran}

\pdfoutput=1

\usepackage{amsmath,amssymb,amsfonts}
\usepackage{algorithmic}
\usepackage{graphicx}
\usepackage{textcomp}
\usepackage{xcolor}
\usepackage{booktabs}
\usepackage[hidelinks]{hyperref}
\usepackage{caption} 
\usepackage{float}
\restylefloat{table}
\usepackage{csquotes}
\usepackage[
    backend=biber,
    bibstyle=ieee,
    citestyle=numeric-comp,
    maxbibnames=2,
    sorting=none
]{biblatex} %
\usepackage{hyperref}

\usepackage{textpos}
\usepackage{framed}
\usepackage{hyperref}

\makeatletter
\newcommand{\linebreakand}{%
  \end{@IEEEauthorhalign}
  \hfill\mbox{}\par
  \mbox{}\hfill\begin{@IEEEauthorhalign}
}
\makeatother

\begin{document}

\title{A `Human-in-the-Loop' approach for\\ Information Extraction from Privacy Policies under Data Scarcity}

\author{\IEEEauthorblockN{Michael Gebauer,
Faraz Maschhur, Nicola Leschke,
Elias Grünewald and
Frank Pallas}
\IEEEauthorblockA{\textit{Information Systems Engineering - TU Berlin} \\
Berlin\\
Email: \{mg, f.maschhur, nl, eg, fp\}@ise.tu-berlin.de}}

\maketitle

\begin{textblock*}{1.1\textwidth}(-1cm, -8cm) %
\begin{center}
\begin{framed}
    \textit{Preprint (2023-05-22) before final copy-editing of an accepted peer-reviewed paper to appear in the\\
    {2023 IEEE European Symposium on Security and Privacy Workshops (EuroS\&PW)}.}\\
    The Version of Record can be found here: \url{https://ieeexplore.ieee.org/xpl/conhome/1820965/all-proceedings}
\end{framed}
\end{center}
\end{textblock*}

\begin{abstract}
Machine-readable representations of privacy policies are door openers for a broad variety of novel privacy-enhancing and, in particular, transparency-enhancing technologies (TETs). In order to generate such representations, transparency information needs to be extracted from written privacy policies. 
However, respective manual annotation and extraction processes are laborious and require expert knowledge. Approaches for fully automated annotation, in turn, 
have so far not succeeded due to overly high error rates in the specific domain of privacy policies. In the end, a lack of properly annotated privacy policies and respective machine-readable representations persists and enduringly hinders the development and establishment of novel technical approaches fostering policy perception and data subject informedness.

In this work, we present a prototype system for a \textit{`Human-in-the-Loop'} approach to privacy policy annotation that integrates ML-generated suggestions and ultimately human annotation decisions. 
We propose an ML-based suggestion system specifically tailored to the constraint of data scarcity prevalent in the domain of privacy policy annotation. 
On this basis, we provide meaningful predictions to users thereby streamlining the annotation process. Additionally, we also evaluate our approach through a prototypical implementation to show that our ML-based extraction approach provides superior performance over other recently used extraction models for legal documents.
\end{abstract}

\section{Introduction} \label{sec:intro}
Machine-readable, structured representations of privacy policies are a door-opener for a broad variety of user-facing privacy-enhancing technologies, ranging from privacy icons \cite{holtz2011privacy, rossi2020} over privacy dashboards \cite{tolsdorf2021case} to interactive privacy chatbots \cite{gruenewaldEnablingVersatile}. Especially when it comes to technically mediated approaches for overcoming the inherent shortcomings of written, legalese privacy policies in matters of comprehensible transparency information (as, e.g., mandated in Art. 12 ff. of the GDPR), the to-be-presented information must be available in a machine-readable and automatically processable form \cite{bonatti_machine_2020, gerl_let_2020, grunewald_2021}. In addition, this machine-readable form must comprise relevant information at a sufficient level of detail and preciseness to actually allow for novel presentation approaches and to preserve legal meaningfulness.

Formal, machine-readable representation languages (e.g., \cite{gerl_lpl_2018, bonatti_machine_2020, grunewald_2021}) play an important part here, but still leave open the gap of transforming pre-existing, textual privacy policies into respective machine-readable representations. As long as data controllers cannot be forced to provide machine-readable privacy policies themselves, performing this transformation is thus one of the most urgent obstacles for alternative -- and, from the perspective of data subjects, often preferable -- policy presentations to actually gain momentum.

Therefore, we propose a \textit{`Human-in-the-Loop'} privacy policy annotation system which facilitates the transition from purely textual privacy policies to machine-readable ones. 
We present an architecture as well as a fully functional prototypical implementation of such a system that enables privacy researchers to continuously annotate privacy policies with required efforts being significantly reduced through suggestions generated with machine learning (ML) and with employed ML models continuously learning new annotations.
To provide high suggestion quality from the very beginning, we employ machine learning algorithms specifically tailored to the additional constraint of \textit{data scarcity}.

To demonstrate the capabilities and the viability of our approach as well as to assess the suitability of different classification approaches and language models for the domain of privacy policies, we furthermore present a benchmark on a consciously small dataset to simulate the situation of \textit{data scarcity}.

Throughout our benchmark, we will focus solely on \textit{data subject rights}, as they are stated in full sentences and do not call for pointy extraction of highly specific facts often codified in single words, which would have been exaggerated for a first evaluation of our overall approach. Given this sentence-focused extraction, the complexity of the task remains, as data subject rights share a lot of semantic similarities with each other. Here, we decided to use \textit{Sentence Bert (SBERT)}, as it enables a significantly more meaningful representation of sentences than other approaches \cite{sbert_reimers_19} and thus is more suited for sentence-based information extraction. Our contributions therefore are:

\begin{itemize}
    \item A novel, so far not pursued approach to semi-automated privacy policy annotation, combining manual expert annotations of privacy policies with ML-generated suggestions in a continuous feedback loop
    \item A fully functional and easy-to-use \textit{information extraction system}, providing an \textit{annotation interface} for users with a continuous learning component and ML-based suggestions for manual annotation 
    \item Benchmarks covering three established information extraction algorithms on a public dataset to test for their capabilities under data scarcity in the specific domain of privacy policies
\end{itemize}

Our respective considerations and findings evolve as follows: In section \ref{sec:background} we give an overview of already existing work in the field and outline the uniqueness of our approach. Section \ref{sec:approach} introduces the basic concepts of our ML-based suggestion engine as well as its model candidates. The implementation of our general approach in the form of a concrete system is presented in section \ref{sec:annotation-system}, whereas section \ref{sec:evaluation} presents an evaluation for all model candidates and outlines the capabilities of our systems. Section \ref{sec:conclusion} concludes.

\section{Background \& Related Work} \label{sec:background}

Automated policy information extraction employing modern methods of natural language processing (NLP) or understanding (NLU) has been a major line of research over the past years \cite{alabduljabbar2021tldr, amarl_2021, wilson-etal-2016-creation, liu2016analyzing, harkous_polisis_2018}.
A well-known framework for automated privacy policy information extraction is Polisis \cite{harkous_polisis_2018}.
Polisis aims to facilitate natural language queries on real-world privacy policies. For this purpose, privacy policies are scraped from the web and divided into segments, which are then classified according to a privacy taxonomy proposed by \cite{wilson-etal-2016-creation} in a machine learning layer. The classification results are then matched with the classes inferred from the natural language query. The machine learning layer used in Polisis comprises privacy-specific static word embeddings, which are then used to classify the segments in a Convolutional Neural Network (CNN)\cite{harkous_polisis_2018}. 

Different from Polisis, however, we do not aim for identifying coarse-grained sections of written privacy policies broadly dedicated to different, rather general subject areas or for respective ``reading hints''. Instead, we want to extract fine-granular and precise transparency information and to thereby transform written privacy policies into semantically rich and analyzable, machine-readable representations. The challenges to be solved and the technical approaches to be considered viable for this purpose strongly differ from those known and used in the Polisis context.

Given the highly domain-specific, legalese language and the complex verbal constructs to be found in privacy policies, \cite{bonatti_machine_2020}, 
fully automated retrieval of privacy policy information for machine-readable formats appears unsuitable as respective results are still severely limited in matters of capability and reliability \cite{wilson-etal-2016-creation, harkous_polisis_2018}.
To a large extent, this limitation can be ascribed to the \textit{scarcity of labeled training data} -- a common and well-known problem in automated privacy policy annotation: Given that training datasets have to be laboriously handcrafted and checked by legal experts, automated policy information extraction is hindered by tremendous efforts in matters of time and cost \cite{amarl_2021, kumar_2020, mousavi2020establishing, wilson-etal-2016-creation}. 

An alternative approach lies in consciously disregarding the goal of completely reliable and accurate policy information extraction in favor of mere \textit{\enquote{considerable hints}} pointing towards sections of a privacy policy presumably denoted to a certain subject or a certain category of information to be provided \cite{alabduljabbar2021tldr, liu2021have, liu2016analyzing, wilson-etal-2016-creation}. However, those approaches rather support -- like the Polisis approach mentioned above -- navigability within written privacy policies and are not eligible for creating detailed and accurate machine-readable representations of privacy policies.
 Additionally, most of these rely on a benchmark dataset introduced by \cite{wilson-etal-2016-creation}, which proposes a privacy taxonomy with ten top level categories that are not aligned with a specific privacy regulation and therefore lack legal expressiveness desirable for such a machine-readable representation. 
Also, such benchmark datasets may cause foreseeable problems such as distributional shifts resulting from changes in privacy policies or well-known limitations when applying benchmark datasets to the real-world \cite{lubbering2022open, amodei2016concrete}.

Therefore, we consciously abstain from these approaches.
Instead, we employ a standardized corpus of machine-readable privacy policies together with their corresponding textual basis, in an \textit{information extraction system}. This system allows us to 
train and extract concrete legal statements, throughout an once-in-a-lifetime training phase.

Regarding machine-readable representation of privacy policy information, a lack of appropriate formats and standardization for representing information in a unified form has previously been identified \cite{alabduljabbar2021tldr}. 
Consequently, standardized languages for representing privacy policy information in machine-readable form are indispensable here. Early such languages include P3P \cite{cranor_p3p_2003, anderson2006comparison} and PPL \cite{angulo2012towards}. However, these are not aligned to a specific privacy regulation and therefore do not include all legally mandatory information. 
To overcome this problem, GDPR-aligned and semantically rich transparency languages have recently been introduced \cite{gerl_lpl_2018, grunewald_2021} and can serve as means for standardizing privacy policy information representation. Given its well-crafted expressiveness and the comprehensive programming toolkit available, we decided to use TILT \cite{grunewald_2021} and the corresponding set of machine-readable policies\footnote{https://github.com/Transparency-Information-Language/tilt-corpus} as our annotated training set.

\section{Approach: Guided Information Retrieval} \label{sec:approach}
\begin{figure}[!ht]
    \centering
    \includegraphics[scale=0.44]{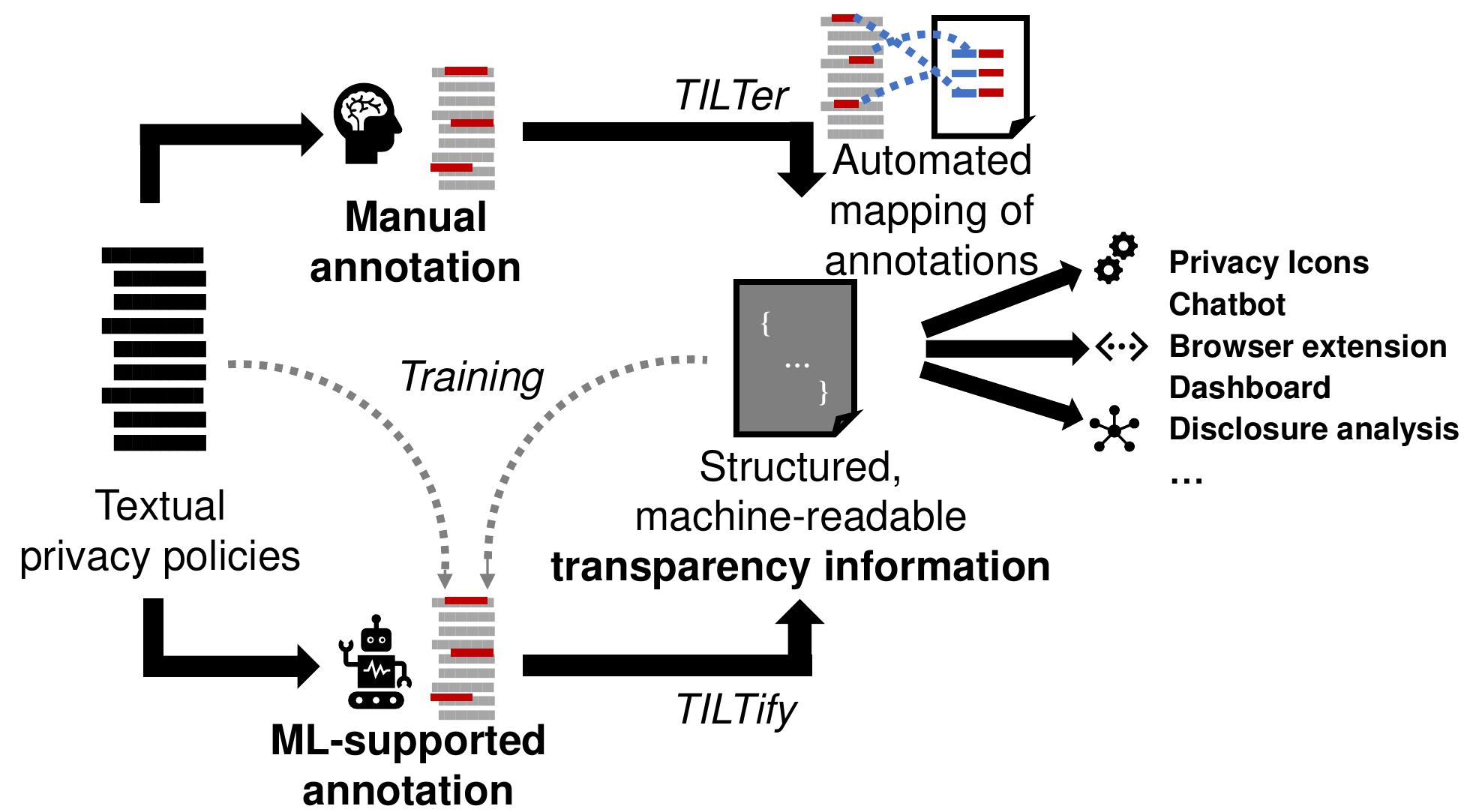}
    \caption{General approach}
    \label{fig:general-approach}
\end{figure}

In light of the above-mentioned considerations, we herein propose an easy-to-use \textit{privacy policy information extraction system} that supports human annotators with ML-based annotation suggestions, as depicted in figure~\ref{fig:general-approach}. Annotators' ultimate decisions are then fed back into continuously (re)training the machine learning model. Through the conscious combination of ML-based suggestions, ultimately human annotation, and a feedback loop to continuously improve suggestion quality, we preserve the preciseness of human privacy policy annotation while continuously decreasing annotation efforts under the conditions of \textit{data scarcity} we face in the domain of privacy policy information retrieval. 

Within this paper, we will focus on the extraction of data subject rights, as they serve as a perfect first example for semi-automated privacy policy information retrieval: On the one hand, data subject rights are typically provided in the form of complete sentences, making the automated extraction less resource-intensive than, for instance, extracting multiple distinct purposes on a word level. On the other hand, different data subject rights are formulated in quite similar form in a given privacy policy, making them non-trivial to correctly extract and distinguish. 
The respective data subject rights covered herein are given in table \ref{tab:rights}.

\begin{table}[H]
\centering
\scalebox{0.8}{
\begin{tabular}{llll}
\toprule
References & Consumer Right \\
\midrule
13(2c) 14(2d)  &     Right to Withdraw Consent \\
13(2b) 14(2c) 15(1e) & Right to Data Portability \\
13(2b) 14(2c) 15(1e) & Right to Correction or Deletion \\
13(2d) 14(2e) 15(1f) & Right to Complaint \\
13(2b) 14(2c) 15(1) & Right to Information \\
\bottomrule
\end{tabular}

\caption{\label{tab:rights} Data subject rights and their respective reference in the GDPR}
}

\end{table}

For the approach, we assume that a privacy policy ($B$) always contains a data subject right $c$, as we want to abstain from completion checks or other approaches (as in, e.g., \cite{bhatia_semantic_2018, torre_ai-assisted_2020}). As data subject rights might be individual sentences in a paragraph, we simplify the approach by retrieving the whole paragraph. Hence, finding the correct paragraph that contains a data subject right $c$ is the core problem of our \textit{Guided Information Retrieval} approach and is achieved by recommending a set of candidate paragraphs $S$ -- that might contain the respective data subject right -- to the user. This also introduces data imbalance, as within one document only a few sentences might contain the respective data subject right. Besides \textit{data scarcity} this makes $c$ hard to retrieve as well.

\subsection{Problem Definition}\label{ssec:problem-def}

  Formalizing the problem further, a privacy policy $B$ can be divided into its respective paragraphs, which we call blobs $b_i$. 
  As policy sections containing data subject rights can span multiple sentences, we use these blobs for our information retrieval process.
  Thus, when trying to extract the data subject right $c$, we always try to retrieve a set of candidate blobs $S$. 
  In case a blob contains a data subject right, we refer to it as $c_i=1$ and call it an annotation. In turn, if it does not, it will be $c_i=0$. Note that this is different from $S$ as $c$ refers to a true data subject right, whereas $S$ is a set of potential candidates for the data subject right in question. Hence, we can formalize our approach as part of a classification problem.

\begin{equation*}
    S = \{b_i : \max_{\forall b_i \in B} P(c_i=1|b_i, \theta) \}
\end{equation*}

Thus, we assume that there exists a perfect retrieval distribution $P(c_i=1|b_i, \theta)$, with true parameter $\theta$ that enables an exact retrieval of a data subject right. In a perfect case, $S$ would thus always contain blobs that actually contain the to-be-annotated data subject right, meaning that $S = \{b_{i} : c_i = 1 \ \forall i\in B\}$. The goal for our ML model, in turn, is to find the perfect retrieval distribution $P(c_i=1|b_i, \theta)$ via an estimator $\hat{f}(b_i, \hat{\theta})$, which we will call extraction model. 

Under \textit{data scarcity}, this estimator $\hat{f}$, trained on a historical dataset, will suffer from poor estimation performance and inevitably make wrong predictions as probabilities are not calibrated well enough. Therefore, we opted for a guided approach, where we rely on user feedback and ultimately human decisions while continuously improving model performance by updating the parameters $\hat{\theta}$ on the basis of these decisions. The feedback provided by users are marked blobs, which are stating whether these particular blobs actually contain a given data subject right or not. These markings are simply called annotations and are generated by users through a dedicated, function-rich annotation interface integrating ML-generated suggestions, full-text search, etc. 

In case an annotator marks a specific data subject right, this translates to $c_i=1$ for a particular blob $b_i$. This way, users continuously generate new annotations that are used for training the estimator and thereby refine retrieval capacities of $\hat{f}$
for data subjects rights.
New annotations for privacy policies other than our policy of concern $B$ are depicted by $\Delta b$. However, as $\hat{f}$ might not be suited to approximate $P(c_i=1|b_i, \theta)$ well enough, and we also have to provide users with feedback from the model through recommendations they are reviewing, we have to refine our information retrieval problem to present the highest $k$ predictions from $\hat{f}$ to the user. Hence, our problem is refined to: 

 \begin{align}
    S = \{b_i : \hat{f}(b_i,\hat{\theta}(\Delta b)) \geq \nu \ \forall b_i \in B \},
    \label{eq:problem}
\end{align}

where $\nu$ is the minimum of the highest $k$ entries in $\hat{f}(b_i,\hat{\theta}(\Delta b))$. As mentioned above, in order to facilitate the feedback loop we will pass on our recommendations $S$ to users, so that they can generate new annotations for feedback $\Delta b$ more quickly.
With this in mind, it becomes obvious that meaningful probability estimates of $\hat{f}$ are vital for a well-functioning feedback loop.
We emphasize that the $k$-based retrieval of blobs within our approach generalizes to many other information retrieval problems.

 \begin{figure*}[t]
\includegraphics[width=0.8\textwidth]{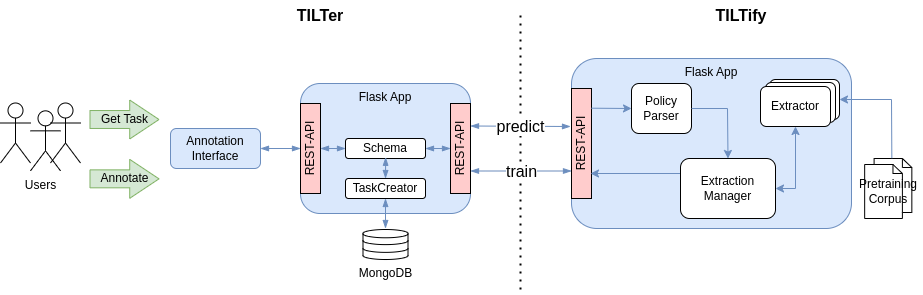}
\centering
\caption{System Architecture of \textit{TILTer} \& \textit{TILTify}, which form the \textit{Information Extraction System}}
\label{graphic:arch}
\end{figure*}

\subsection{Extraction Model Options}\label{ssec:extraction-models}

The problem of finding an $\hat{f}$ that appropriately captures $P(c_i=1|b_i, \theta)$ becomes particularly challenging under \textit{data scarcity}. Therefore, we decided to implement and investigate multiple model candidates for $\hat{f}$. 

\subsubsection*{Static Word Embeddings \& Naive Bayes}
Static word embeddings have a long history, starting with One-Hot encoded vectors, \textit{TF-IDF} \cite{tf_idf} and also \textit{Latent Semantic Analysis} \cite{landauer1998lsa}. Deep Learning based Embeddings first used neural networks to create meaningful vector representations of words, such as \textit{Word2Vec} \cite{mikolov_13}. \textit{Spacy} offers one of the so-called static word embedding techniques, namely \textit{Token2Vec}\footnote{\url{https://spacy.io/api/tok2vec}}. \textit{Token2Vec} employs a Convolutional Neural Network, together with Hash Embedding of sub-features of individual words.
These embeddings suffer from a lack of context awareness, as static embeddings are context-independent and do not account for the context a singular word appears in \cite{elmo_18}. For our purposes, we use these embeddings and pass them on to a \textit{Gaussian Naive Bayes Classifier}. Herein, every dimension of the embedding is assumed to follow a gaussian distribution. These dimensions are assumed to be independent of each other, as a \textit{Gaussian Naive Bayes Classifier} links all dimensions solely to the target variable. The target variable in our case is a label on whether a specific blob contains a specific data subject right or not. The static embeddings produced by the above-stated model represent singular blobs within the privacy policy.

\subsubsection*{BinaryBERT}
\textit{BERT} \cite{bert_2018} is a transformer-based language model, which is capable of solving multiple NLP tasks. It is divided into two major steps: pre-training and fine-tuning. The pre-training involves a Masked Language Model approach, where 15 percent of all words are omitted and then predicted based on their context, and Next Sentence Prediction, where the model has to identify sentences that are following after another, from random ones. After this pre-training phase, \textit{BERT} can be fine-tuned for specific tasks, like Named-Entity Recognition, Question Answering, Sequence Tagging and many more. 
As BERT showed superior performance over many NLP tasks, also for text classification \cite{bert_2018, sun2019fine} and showed promising performance in other privacy policy extraction approaches \cite{liu2021have}, we decided to utilize it as well.
For our purposes, we decided to employ the implementation of huggingface, which provides a \textit{BERT} model specifically for sequence classification\footnote{\url{https://huggingface.co/docs/transformers/model_doc/bert\#transformers.BertForSequenceClassification}}. We also use a pre-trained \textit{BERT} model, which we fine-tune on our data subject rights identification task. We pass blobs to the \textit{BERT}-model and provide training labels, which indicate whether a blob makes a statement about a specific data subject right or not. One \textit{BERT} model is only suited for one particular data subject right, making it a binary classification task. We therefor call this mode candidate \textit{Binary Bert (BBERT)} and use Binary Cross Entropy as a loss function for optimization. As an optimizer, we use Adam with Decoupled Weight Decay Regularization \cite{adamw_2017}.

\subsubsection*{SentenceBERT}
\textit{SBERT} is an addition to \textit{BERT}. As \textit{BERT} is not well suited for sentence classification or information retrieval \cite{sbert_reimers_19}, \textit{SBERT} tries to overcome this by utilizing a siamese network and a collection of losses. These losses aim to tune sentence embeddings based on BERT for a stronger semantic distinction, which arises from the supplied training corpus. In our approach, we utilized the Triplet Loss Function, as in equation \ref{Triplet Loss Function} \cite{sbert_reimers_19}.

\begin{equation}
\label{Triplet Loss Function}
    J(q, b_p, b_n) = max(||q - b_i||_2 - ||q - b_j||_2 + \epsilon, 0),
\end{equation}

where $i$ contains the specific data subject right and $j$ does not.
For our purposes, we set $q$ to the legal texts about data subject rights stated in the GDPR articles in table \ref{tab:rights} and matched these to the data subject right mentioned in the privacy policies. This had two major advantages over the other two approaches and the other loss options provided in \textit{SBERT}. Firstly, we could increase the size of the training data respectively, as we had to pass all possible combinations of fitting and unfitting blobs into the model and secondly, by $\epsilon$ we had direct control over the distinction between fitting or unfitting blobs within a policy.
Additionally, this approach also reduces the severity of data imbalance due to the Triplet Loss Function.

\section{Annotation System} \label{sec:annotation-system}
The \textit{information extraction system} is divided into two major parts -- \textit{TILTer}\footnote{\url{https://github.com/DaSKITA/tilter}} and \textit{TILTify}\footnote{\url{https://github.com/DaSKITA/tiltify}} -- and represents the prototypical implementation of our \textit{Guided Information Retrieval} approach. Of these, the former one is responsible for communication with the user through an \textit{annotation interface} and the persistence of users' annotations. The latter is the learning and inference module, which trains on persisted user annotations and provides suggestions when processing respective textual privacy policies. In figure \ref{graphic:arch}, an overview of the whole system architecture is presented. In particular, we receive user inputs via the \textit{annotation interface}, through which human annotators can navigate to different annotation tasks and annotate the respective transparency information labels (TILT labels). These annotations are then processed in a backend, powered by a Python Flask app, and saved in a MongoDB. 

Depending on the users' inputs, the \textit{TILTer} sends content to different endpoints of \textit{TILTify}, which either trigger a training or inference procedure within \textit{TILTify}. All content received in \textit{TILTify} goes through a parser first, as we send privacy policies with their respective annotations to \textit{TILTify} as JSON strings. The policy parser converts them into a \textit{Document} object, which is used within \textit{TILTify} for internal communication between components. Inference and training within \textit{TILTify} is orchestrated through the \textit{Extraction Manager} that is responsible for triggering routines within the respective \textit{Extractors}. When starting the system, initially all \textit{Extractors} are trained on a small dataset, which provides some annotations for \textit{data subject rights}. Outputs received from every \textit{Extractor} are ultimately propagated back to the \textit{TILTer} and then presented to the user as considerable hints. This loop can then be continuously repeated and even executed in parallel by multiple human-in-the-loop annotators for guided information retrieval.

\subsection{Annotating Privacy Policies}

All privacy statements to be annotated are fed into \textit{TILTer} through API calls, saved into the application database and exposed to users through the \textit{Annotation Interface}. Subsequent user interaction is conducted directly in the \textit{Annotation Interface} as well. Annotations applied in there pass through the \textit{TILTer} backend, are saved and forwarded to \textit{TILTify}, where they may invoke training processes. The candidates offered for annotation are specified in a schema file in \textit{TILTer}. This schema file enables \textit{TILTer} to be used for the annotation of an entire TILT document. Each candidate offered for annotation is thus an element in a TILT-document, formerly introduced as TILT-labels.

\subsubsection*{Annotation Interface}
\begin{figure}[ht]
\includegraphics[width=1\linewidth]{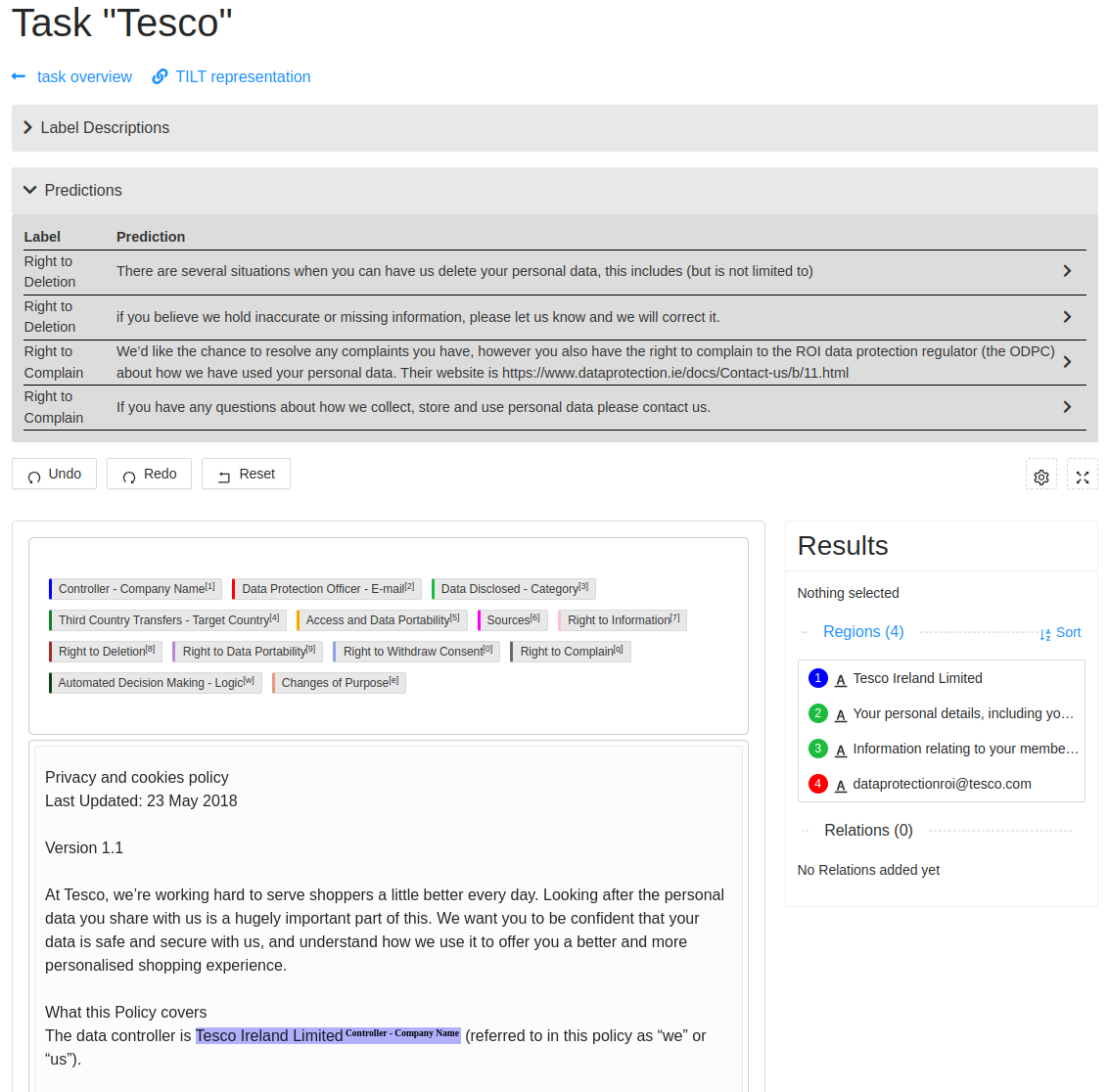}
\centering
\caption{\textit{TILTer Annotation Interface}}
\small
\label{graphic:tilter}
\end{figure}

Figure \ref{graphic:tilter} shows the \textit{Annotation Interface}, which enables well-informed annotation of privacy policies using a pre-defined set of TILT-labels obtained from the schema file\footnote{This schema file follows the available building blocks of the TILT schema.}. Predictions obtained from models provided by \textit{TILTify} are offered to support the user and thus are the formerly defined set of suggestions $S$. All TILT-labels offered for the annotation are explained in more detail under the retractable \textit{Label Description} table. A second retractable table encloses the \textit{TILTify} predictions, through which we enable easy navigation to the marked blob.
Below the tables on the left, the available TILT-labels are shown while and on the right, current annotations are displayed. The policy text is presented below these TILT-labels. Annotations are made by selecting a label and marking the passage to be annotated. The TILT-label overview continues to be shown when scrolling.
The annotation functionalities of the \textit{Annotation Interface} are based on Label Studio Frontend\footnote{\url{https://github.com/heartexlabs/label-studio-frontend}}.

\subsubsection*{Task Creation}
The \textit{TILTer} backend allows for the processing of privacy policies through REST API calls, which create a task containing the policy text as well as a set of TILT-labels. Since TILT labels are ordered hierarchically, annotating certain blobs leads to the creation of new tasks on the same policy, with TILT-labels from the next hierarchical level.
This way, we ensure the user is not overwhelmed by the numerous annotation options through numerous TILT-labels.

\subsection{Learning \& Inference}

As previously stated, \textit{TILTify} is responsible for learning and inference within our system. It either generates annotation suggestions $S$ for users or uses their inputs $\Delta b$ to improve the model performance. The entire communication within \textit{TILTify} happens in the form of a \textit{Document} class, which represents the respective data and follows the notation we introduced in section~\ref{ssec:problem-def} for blobs and documents. We designed \textit{TILTify} to be extensible for other types of documents as well, while making it compatible to changing legal requirements.

\subsubsection*{Policy Parser}

The policy parser is responsible for converting a privacy policy provided in the form of a JSON string into the \textit{Document} class. A \textit{Document} class wraps a list of blobs, which contain paragraphs within a privacy policy, thus representing $b_i \in B$. A blob contains the text of a paragraph and annotation objects, that are either provided by users or predicted by models, similar to the annotation we introduced earlier $c_{i}$. So far, we do not provide a token-based annotation class, as we only focus on \textit{data subject rights}. However, we are aware that TILT will at a later stage also require annotations on a token level as well (e.g., controller name, third-country transfers, etc.).

\subsubsection*{Extractor}
The \textit{Extractor} provides an object for standardized communication between the Extraction Manager and the extraction routines and models $\hat{f}$. Thus, an \textit{Extractor} maps a provided TILT-label to an extraction model. The label represents the content to be extracted from the \textit{Document} object, which is a subset of all annotation objects within a \textit{Document} object, that matches the label, as there are more than one annotation within a \textit{Document}. As described above, we only provide sentence-based annotations so far, hence \textit{Extractors} also focus on sentence-based information retrieval. 
Annotations within a \textit{Document} object are either inferred or used for training by extraction models. Therefore, extraction models wrap models of an existing ML library (e.g., \textit{huggingface}, \textit{scitkit-learn}, \textit{pytorch}), to define their prediction, training, loading, and saving routines, as well as tying them to a data preprocessor. Data preprocessors are creating the data format the model needs for inference and training out of the \textit{Document} objects provided together with their annotations.
When a prediction is performed by an extraction model, writing predicted content into the respective \textit{Document} object is done by the \textit{Extractor} as well and is then propagated to the extraction manager.

\subsubsection*{Extraction Manager}

The Extraction Manager orchestrates \textit{Extractors}, depending on the inputs received over the REST API. The extraction manager contains a model- and an extractor registry, where it contains all possible extraction models (see \ref{ssec:extraction-models}) and all existing \textit{Extractors} that are created out of the existing extraction models and a provided label. Orchestration is done via interaction with the extractor registry within the extraction manager, making it thus the central point of access for training, prediction, loading, saving, and initialization of all \textit{Extractors}. It also passes on the results from \textit{Extractors} to the REST API, so that they can be communicated to \textit{TILTer}.

\section{Evaluation} \label{sec:evaluation}
In order to prove the validity of our approach, we conducted multiple experiments. We performed information retrieval tasks for data subject rights over a formerly annotated dataset and %
checked for the validity of model estimates to evaluate the actual applicability of the different extraction model options under data scarcity.

\subsubsection*{Dataset}

We created a public dataset\footnote{\url{https://github.com/DaSKITA/tiltify/tree/main/data/annotated_policies}}, consisting of 60 annotated policies according to the TILT format in German language. Each of these contains differing amounts of data subject rights, given real-world privacy policies' significant variance in this regard. The dataset can be used by other research to measure against our benchmarks. Again resembling real-world givens, the data exhibits a reasonable imbalance: in total, there are 16635 blobs, whereas they only contain a few data subject rights. Table \ref{tab:results_all} illustrates the number of blobs for every data subject right in the last column. Some data subject rights have multiple annotations within one document. In such cases, we determine the information retrieval to be successful as soon as one of the entries is retrieved by $\hat{f}$. Data scarcity manifests in the fact that we only used 60 documents for training and testing.

The dataset was created through iterative cycles of annotation rounds. In the first round, student researchers annotated the data, followed by a validation performed in the second round by more senior members of the project team. Flaws in annotations in the first round could thus be removed, ensuring more validity in provided annotations.

\subsubsection*{Experimental Setting}

\begin{table}[H]
\resizebox{\columnwidth}{!}{
\begin{tabular}{llll}
\toprule
 & BinaryBERT & SentenceBERT & GaussianNB \\
\midrule
$5$-rank                  &     0.08 (0.12) &    0.90 (0.00) &       0.00 (0.00) \\
$10$-rank                 &     0.08 (0.12) &    0.90 (0.00) &      0.17 (0.00) \\
$25$-rank                 &     0.31 (0.00) &  0.93 (0.04) &      0.31 (0.00) \\
classification &       0.25 (0.35) &  0.58 (0.02) &      0.32 (0.00) \\
support                   &      11  &   11&      11 \\
\bottomrule
\end{tabular}

\caption{\label{tab:results} F1-Scores for the \textit{Right to Withdraw Consent}.}
}
\end{table}

\begin{table*}[h]
\centering
\scalebox{0.9}{
\begin{tabular}{lllll}
\toprule
  & BinaryBERT  & SentenceBERT  & GaussianNB & \#Blobs \\
\midrule
Right to Information      &     0.00 (0.00)  &  \textbf{0.93 (0.05)} &     0.18 (0.08)  &  83 \\
Right to Deletion         &     0.00 (0.00) &  \textbf{0.86 (0.05)}  &       0.00 (0.00) & 87 \\
Right to Data Portability &       0.00 (0.00)  &  \textbf{0.86 (0.05)} &       0.00 (0.00) & 77 \\
Right to Withdraw Consent &     0.08 (0.12)  &    \textbf{0.90 (0.00)} &       0.00 (0.00) & 95 \\
Right to Complain         &      0.15 (0.00) &       \textbf{0.93 (0.03)} & 0.29 (0.00) & 80 \\

\bottomrule
\end{tabular}

}
\centering
\caption{\label{tab:results_all} $5$-Rank F1-Scores for all models over all data subject rights.}
\end{table*}

For all models, we decided to use the same training and test data. Each of the models was trained for two repetitions over that same split, to account for different model initializations of parameters. Thus we are ruling out as many random effects as possible to maintain adequate comparability. Hence we always report mean values for classification metrics and the respective standard deviation. Training epochs for the \textit{Gaussian Naive Bayes} model were set to 100, while \textit{SBERT} and \textit{BBERT} only use 5 training epochs, as they are fine-tuned for the task of data subject right retrieval. For all models, we use the default hyperparameter values, except for \textit{BBERT} as we had to define an optimizer, where we choose Adam, with Decoupled Weight Decay Regularization \cite{adamw_2017} with learning rate $1\cdot10^{-3}$ and weight decay of $1 \cdot 10^{-5}$. As the dataset has a high data imbalance, we trained all models with a weighted sampling approach, where training examples were down- and upsampled to achieve an equal amount for paragraphs with ($c_i=1$) and without a data subject right ($c_i=0$). As we use German privacy policies for our experiment, we used a German version of \textit{BERT}\footnote{\url{https://huggingface.co/dbmdz/bert-base-german-cased}} to obtain meaningful representations in \textit{BBERT} and \textit{SBERT}.

\subsubsection*{$k$-Rank Metrics \& Classification}

The $k$-rank metrics describe the highest $k$ model outputs, whereas an information retrieval with this approach was considered to be successful if among the $k$ outputs of $\hat{f}$ a real data subject right $c_i=1$ can be found (thus, when the correct paragraph was among those suggested to an annotator).
The resulting F1-Scores in table \ref{tab:results_all} are clearly in favor of \textit{SBERT}, as it outperforms the other two models in all metrics for all data subject rights. The surprisingly strong performance of the model might be attributed to its better-suited sentence embeddings and stronger capabilities of manipulating them, due to the Triplet Loss Function. It is also obvious that some data subject rights are harder to classify than others. The \textit{Right to Data Portability} and the \textit{Right to Deletion} seem to be quite hard to identify for some models. Despite being more complex, \textit{BBERT} does not perform significantly better than the \textit{Gaussian Naive Bayes} model. This might be attributed to the data-hungry nature of \textit{BERT} and the tendency of non-deep learning models to perform good also on fewer data. Nevertheless, the static word embeddings used in \textit{Gaussian Naive Bayes} might contribute to its poor performance, as the embeddings might not be able to account for the complexity of the task. Both \textit{Gaussian Naive Bayes} and \textit{BBERT}  seem to strongly suffer from data scarcity, as they do not provide any meaningful extractions. For this case, it remains obvious that \textit{SBERT} is the candidate of choice when confronted with data scarcity as prevalent in the domain of privacy policy information extraction.

In table \ref{tab:results}, the $k$-rank metrics also underline that \textit{SBERT} is way better at making suggestions than the other two models: Using the example of the \textit{Right to Withdraw Consent}, the first $k$ suggestions of \textit{SBERT} actually contain a fitting paragraph way more often than those provided by the others. Even when increasing the number $k$ of estimated paragraphs to consider, the other two models are not able to catch up. However, looking at the \textit{classification} performance of all models, \textit{SBERT} does not seem to be remarkably better. Hence, it might not be superior in the exact identification of data subject rights, but rather seems to give better probability estimates for data, which are more useful in our human-in-the-loop approach.

\subsubsection*{Calibration}

Stronger semantic expressiveness for sentences also becomes obvious when we look at the calibration of the models. For this purpose, we will make use of the Brier-Score \cite{lubbering2022bounding}, which is defined as

\begin{equation}
    BS = \frac{1}{N}\sum_{i=1}^{N}(c_{i} - \hat{f}(b_i,\hat{\theta}))^2
\end{equation}

and calculates the squared error between the model estimates and the real label ($c_{i}$), given a blob $b_i$, over all existing blobs in the dataset $N$. Hence, the better calibrated a model, the lower is the Brier Score. A well-calibrated model gives more-aligned probability estimates, as it is more proficient in expressing uncertainty about its predictions due to its estimates. Thus well-calibrated models should be more proficient in a $k$-ranking approach. 

In table \ref{tab:brier} we can see the Brier scores for our respective models for every data subject right. Herein we choose to adjust the Brier Score with sample weights so that the data imbalance mentioned earlier is represented in the score and adjusted for the few occurrences of data subject rights. Throughout all data subject rights, \textit{SBERT} has far lower Brier Scores, which indicate a better calibration than the other models. This shows that the semantics of the embeddings created by \textit{SBERT} seem to express data subject rights way better, which is in line with the findings of \cite{sbert_reimers_19}. Hence, despite data scarcity, \textit{SBERT} gives well-adjusted semantic representations, which makes it far more fitting for information retrieval under data scarcity than other models. However, \textit{BBERT} still seems to be better calibrated than the \textit{Gaussian Naive Bayes}. Still, both are rather incapable of successfully extracting data subject rights.

\begin{table}[h]
\resizebox{\columnwidth}{!}{
\begin{tabular}{llll}
\toprule
{} & BinaryBERT & SentenceBERT & GaussianNB \\
\midrule
Right to Information      &      0.25 (0.00) &   \textbf{0.03 (0.00)} &      0.51 (0.00) \\
Right to Deletion         &      0.25 (0.00) &  \textbf{0.05 (0.01)} &      0.48 (0.00) \\
Right to Data Portability &      0.24 (0.00) &   \textbf{0.03 (0.00)} &      0.46 (0.00) \\
Right to Withdraw Consent &     0.25 (0.01) &   \textbf{0.04 (0.00)} &      0.53 (0.00) \\
Right to Complain         &     0.26 (0.02) &   \textbf{0.01 (0.00)} &      0.48 (0.00) \\
\bottomrule
\end{tabular}

\caption{\label{tab:brier} Brier-Scores for all data subject rights.}
}
\end{table}

\section{Conclusion} \label{sec:conclusion}
In this paper, we propose a ML-assisted \textit{Guided Information Retrieval} approach to streamline the process of privacy policy annotation. We design and prototypically implement a respective \textit{Annotation System} which facilitates continuous privacy policy annotation in a "\textit{human-in-the-loop}" manner supported by ML-generated suggestions. 
Using the example of data subject rights, we showed that using \textit{SBERT} to generate respective suggestions is far superior to any other established models for the extraction of information out of privacy policies, particularly in $k$-based retrieval scenarios like the one employed for our \textit{Guided Information Retrieval} approach. Insofar, it is surprising that \textit{SBERT} has so far not broadly been used in the field of (semi-) automated privacy policy annotation. 

Our prototypical \textit{Annotation System} can of course be improved further, especially by introducing a more production-ready \textit{TILTify} architecture with technologies like Apache Spark\footnote{\url{https://spark.apache.org/docs/latest/}}, ML-Flow\footnote{\url{https://mlflow.org/docs/latest/index.html}}, or other MLOps Frameworks, which can take the serving of the \textit{Extractors} to the next level and thus open it for more users. Additionally, relevant legal information to be extracted from privacy policies is not limited to sentences or paragraphs but also includes single words (tokens), which might be worth retrieving. Thus, investigating the predictive power and validity of our approach for word-based extraction is one of the next steps to take. We are also aware, that the retrieval of highly specific facts on a word level, may call for more complex and adjusted models, making model choice dependent on the legal facts one wants to extract. Lastly, continuous re-training of the models based on human annotations should successively increase suggestion quality further. However, faulty annotations by human error, may also introduce bias into trained models. Thus continuous testing of models on a provided test dataset maybe required, while replacement of a new model version should be guided by performance measures on the respective test dataset.

As a first prototype, however, our system already proves our point that \textit{Guided Information Retrieval} via \textit{SBERT} can significantly accelerate the extraction of legal information from privacy policies. Further research in this direction should -- at least in the medium term -- pave the way for an easier and economically more feasible transfer of written, legalese privacy policies into machine-readable representations. In the end, this will also spur and accelerate the development of transparency-enhancing technologies for presenting legally relevant information to data subjects in novel, technically mediated, and more comprehensible ways.

\subsubsection*{Acknowledgements}
\begin{small}
The work behind this paper was partially conducted within the project DaSKITA, supported under grant no. 28V2307A19 by funds of the Federal Ministry for the Environment, Nature Conservation, Nuclear Safety and Consumer Protection (BMUV) based on a decision of the Parliament of the Federal Republic of Germany via the Federal Office for Agriculture and Food (BLE) under the innovation support program.
\end{small}

\renewcommand*{\bibfont}{\footnotesize}
\printbibliography

\end{document}